\documentclass{ieeeaccess}
\usepackage{cite}
\usepackage{amsmath,amssymb,amsfonts}
\usepackage{algorithmic}
\usepackage{graphicx}
\usepackage{textcomp}
\usepackage{color}
\usepackage{soul}

\usepackage{refcount} 
\usepackage{cases}
\usepackage{alltt}
\usepackage{hyperref}

\allowdisplaybreaks

\begin{document}
\history{Final,  18 February, 2021}
\doi{DOI to be assigned}
\title{Comment on ``Dr.~Bertlmann's Socks in a Quaternionic World of Ambidextral Reality''}
\author{\uppercase{Richard D.~Gill}\authorrefmark{1}}
\address[1]{Mathematical Institute, Leiden University, Netherlands (e-mail: gill@math.leidenuniv.nl)}
\tfootnote{No financial support was obtained for this research}

\markboth
{Richard D. Gill: Comment on ``Dr.~Bertlmann's socks...''}
{Richard D. Gill: Comment on ``Dr.~Bertlmann's socks...''}

\date{18 February, 2021; final version}                                           

\begin{abstract}
\noindent I point out critical errors in the paper ``Dr.~Bertlmann's Socks in a Quaternionic World of Ambidextral Reality'' by J. Christian, published in {\em  IEEE Access}. Christian's model does not generate the singlet correlations but in fact simply reproduces the Bertlmann effect. John Bell's colleague Reinhold Bertlmann of CERN, in his younger days, always wore one pink and one blue sock, at random. The moment you saw his left foot, you knew what colour sock would be on his right foot. Action at a distance? As John Bell liked to explain, quantum entanglement cannot be explained away in such an easy way. Yet Christian's model assigns the two particles of the EPR-B experiment an equal and opposite spin at the source, the choice being determined by a fair coin toss. However they are measured, these spins are recovered. Christian's computer simulation works by not actually simulating his model at all but by almost directly tracing the negative cosine built into his computer algebra package. Bell's theorem has not been disproved. Debate as to what it means for the foundations of physics as well as for quantum information engineering (quantum communication, computation) is more lively today than ever before. A possible role for Geometric Algebra is still wide open and deserves further investigation, informed by a proper understanding of the mathematical content of Bell's theorem.
\end{abstract}


\titlepgskip=-15pt

\maketitle

\section{Introduction}

\noindent  Bell's 1964 theorem \cite{Bell} states that the conventional framework of quantum mechanics is incompatible with a physical principle called \emph{local realism}. Bell's theorem is a cornerstone of modern quantum information theory, and of quantum computing. Proof that it is wrong would unleash a revolution in science with enormous impact on society and technology. Every textbook on quantum mechanics would have to be rewritten.

At the core of Bell's proof of his theorem is an elegant and simple probability inequality, going back to Boole (1853) \cite{Boole}. However, Christian \cite{IEEEAccess2} claims in a recent paper in \emph{IEEE Access} to have a counterexample. His other recent papers \cite{RSOS}  and \cite{IEEEAccess1}, the last of which was also published in \emph{IEEE Access}, make the same claim. They have the common purpose of disproving Bell's theorem by giving a local realistic model, which in a nutshell means a classical physical model, which reproduces quantum correlations. Christian's models are formulated in the language of geometric algebra (GA), see Doran and Lasenby (2003) \cite{DoranLasenby}.

In this paper, I will go straight to the heart of Christian's new paper \cite{IEEEAccess2}. I show that the mathematical definitions in the new paper lead to a trivial model. The author's calculations within his model are incorrect.  Inconsistent and ambiguous notation distracts, but cannot hide elementary errors in reasoning and mathematics. Bell's theorem is not contradicted. The computer simulation reported in the paper is not a simulation of the model described in the paper, and it does not prove anything. 
 
Christian's idea that quantum correlations are explained by the geometry of space might seem appealing, but his work lends no support to this idea. Anyway, such an explanation would not be ``local'' in any meaningful sense. Christian seems to see the local spatial coordinate system of Alice being the mirror image of Bob's, the two orientations being determined completely at random, again and again! However, in modern accounts of Bell's theorem, angles and orientations play no role whatsoever. The new generation of loophole-free Bell experiments \cite{Hensen2015} measure correlations between four binary variables: two binary inputs and two binary outputs; one input and output at each of two distant locations. Bell does not take account of the geometry of space because his argument, on the side of local realism, does not depend on it in any way whatsoever.

\section{The heart of the matter}

In the \emph{language of probability theory}, the mathematical core of Bell's original proof of his theorem is the assertion that one cannot find a single probability space on which are defined random variables $X_{\mathbf a}$ and $Y_{\mathbf b}$ taking values in the set $\{-1, +1\}$, for all ${\mathbf a}$, ${\mathbf b}$, unit vectors in $\mathbb R^3$,  and such that
$$
{\mathbb E}(X_{\mathbf a} Y_{\mathbf b}) ~=~ -{\mathbf a} \cdot {\mathbf b}
\eqno(1)
$$
for all $\mathbf a$, $\mathbf b$. 
Moreover, the expectation values of $X_{\mathbf a}$ and $Y_{\mathbf b}$ are all zero. The point is that these statistical properties are predicted in a particular set-up studied in quantum mechanics called the EPR-B model after the famous papers of Einstein, Podolsky and Rosen, 1935 \cite{EPR}, and Bohm and Aharonov, 1957 \cite{BA}. This collection of joint probability distributions of  two binary variables, indexed by pairs of directions $({\mathbf a}, {\mathbf b})$, is called ``the singlet correlations''.

Both in real experiments and in quantum mechanical theory, one can only look at one pair of directions at a time. Thus, one can perform an experiment and observe one realisation of a pair of binary variables $(X_{\mathbf a}, Y_{\mathbf b})$ for a given pair of settings $({\mathbf a}, {\mathbf b})$. This will then be repeated many times, for the same or for different setting pairs. Notice that when $\mathbf a =\mathbf b$, QM predicts perfect anti-correlation. In the EPR-B thought experiment, we imagine performing a measurement on each of two spin-half particles at a great distance from one another. The result $X_{\mathbf a}$ of measuring Alice's particle is obtained before the direction $\mathbf b$ in which Bob has chosen to measure his could possibly be known at Alice's location. Bob's setting can have no effect whatsoever on Alice's outcome. But Bob could measure in any direction, and if Alice were to measure in that same direction, her outcome would be the opposite of Bob's. This suggests that all the outcomes $X_{\mathbf a}$, $Y_{\mathbf b}$ for all possible directions $\mathbf a$, $\mathbf b$ exist in advance, perhaps as deterministic functions of the chosen directions and of some \emph{hidden variables}.

Classical physical thinking therefore suggests that a \emph{local hidden variables} (LHV) model could reproduce the singlet correlations. It would consist of two functions $\mathcal A(\mathbf a, \lambda)$ and  $\mathcal B(\mathbf b, \lambda)$, taking the values $\pm1$, and a probability distribution $\rho$ over the space $\Lambda$ of possible values of the hidden variable $\lambda$. Now simply change notation: write $\omega$ instead of $\lambda$, $\Omega$ instead of $\Lambda$. The random variable $A_{\mathbf a}$ is just the function $\mathcal A(\mathbf a, \cdot)$ defined on $\Omega$. The word ``local'' refers to the fact that each of those measurement functions depends only on the measurement setting at the relevant location. The hidden variable need not be thought to be ``localised'' at any particular place. It could contain components to account for randomness at each measurement location in interaction with the measurement setting chosen there. Moreover, all these contributions can be correlated with one another. 

Bell (1964) proved that there can be no LHV model which even approximately reproduces the singlet correlations. His proof was soon improved by Clauser, Horne, Shimony, and Holt (1969) \cite{CHSH}. Consider two directions $\mathbf a_1$ and $\mathbf a_2$, and two directions $\mathbf b_1$ and $\mathbf b_2$, all in the equatorial plane. They correspond to angles in $[0, 2\pi)$. Consider angles $\alpha_1 = 0$, $\alpha_2 =\pi/2$; $\beta_1 = \pi/4$, and $\beta_2 = 3\pi/4$. Thus $-\mathbf a_i \cdot \mathbf b_j = -\cos(\alpha_i - \beta_j)$.  Three of the differences $\alpha_i - \beta_j$ are equal to $\pm \pi/4$ and one is $\pm 3 \pi/4$. This means that $-\mathbf a_i\cdot \mathbf b_j$ equals $-\cos(\pi/4)= - \sqrt 2 / 2$ for three of the four combinations and $+\cos(\pi/4) = +\sqrt 2 /2$ for the fourth. I will temporarily abuse notation and write $X_i$, $Y_j$ for $X_{\mathbf a_i}$,  $Y_{\mathbf b_j}$, $i,j = 1, 2$. Consider the expression $X_1Y_2 -X_1Y_1 - X_2 Y_1 - X_2 Y_2 = X_1(Y_2 - Y_1) - X_2(Y_2 +Y_1)$. Since the four random variables take the values $\pm 1$ only, one of the terms in brackets equal $\pm 2$ and the other equals $0$. They are multiplied by $\pm 1$ and subtracted. The whole expression therefore takes the value $\pm 2$. Its mean value therefore cannot exceed $+2$. Thus we obtain the one-sided Bell-CHSH inequality 
$$
\mathbb E X_1 Y_2 - \mathbb E X_1 Y_1 - \mathbb E X_2 Y_1 - \mathbb E X_2 Y_2 \le 2.
\eqno(2)
$$
However, if the joint probability distribution of these four random variables would reproduce the singlet correlations, the same expression would have to equal $2 \sqrt 2$.

The mathematical core of Bell's proof of his theorem can also be expressed in the \emph{language of distributed computing}. It then becomes the assertion that one cannot write programs for two separated classical computers , each receiving as inputs streams of directions and generating as outcomes streams of numbers $\pm 1$, such that the correlation between the outputs given the inputs ${\mathbf a}$, ${\mathbf b}$ is $ -{\mathbf a} \cdot {\mathbf b}$. In the appendix to this paper we explain the equivalence of the probability theory no-go theorem, and the distributed computing no-go theorem. 

Christian's latest paper \cite{IEEEAccess2} includes what appears to be a construction of a LHV model with measurement functions $\mathcal A$ and $\mathcal B$ having the prescribed properties. His hidden variable is a fair coin toss, and he gives it a physical interpretation as left-or right-handedness of a coordinate system. He also supplies a computer simulation. In this note we will show that Christian's construction (just as in his preceding works) actually leads to the unwanted result
$$
\mathbb E(X_{\mathbf a} Y_{\mathbf b}) ~=~ -1
\eqno(3)
$$
for all ${\mathbf a}$, ${\mathbf b}$. One could call this the Bertlmann's socks correlation. Christian's model is a local hidden variables model, a trivial one. It does not reproduce the predictions of quantum mechanics for the EPR-B thought experiment. Christian has not disproved Bell's theorem. 

We will also explain what is wrong with his computer simulation. It apparently does reproduce the singlet correlations, but this means that it cannot actually be a simulation of Christian's model. By Bell's theorem, it cannot even be a simulation of a local hidden variables model. Indeed, inspection of the code shows that the program merely computes $-\mathbf a \cdot \mathbf b$ plus some random bivector noise of mean zero, directly from $\mathbf a$ and $\mathbf b$.

Heine Rasmussen (in an internet forum debate) pointed out the following short cut to realising that Christian's claims must be false. For $\mathbf a \ne \pm \mathbf b$, the probability distribution of the pair of binary variables $(X_{\mathbf a}, Y_{\mathbf b})$ predicted by quantum mechanics gives positive probability to each of four distinct joint outcomes $(\pm1, \pm1)$. There is no way one can simulate a single draw from a probability distribution over four outcomes, each  of positive probabilty, as a deterministic function of the outcome of \emph{one} fair coin toss. Christian's hidden variable $\lambda$, which one may identify with the elementary outcome $\omega$ of the alleged probability model on which all those random variables are defined, is a fair coin toss, and in his model, the results of measurement of spin of the two particles in any two directions are functions only of $\lambda$ and of the relevant direction. Christian's computer simulation program uses a fair coin toss to average Geometric Algebra products using the fundamental GA formula $\mathbf a\cdot \mathbf b =\frac12 \mathbf a \mathbf b + \frac12 \mathbf b \mathbf a$.

For those needing introduction to the whole field of Bell's theorem, Bell's inequalities, local hidden variables, loophole-free Bell experiments, and computer simulations thereof, the appendix to this paper supplies some further background and make some further remarks concerning issues raised by the referees.

\section{The geometric algebra and the computer simulation}

I will now go into specific problems in Christian's newest paper \cite{IEEEAccess2}, which the reader will need to have to hand. Christian formulates his model in terms of the Clifford Algebra $\text{Cl}_{(3, 0)}(\mathbb R)$. Recall that the algebra is generated by starting with three elements (called \emph{basis vectors}) $\mathbf e_1$, $\mathbf e_2$, and $\mathbf e_3$, which anticommute with one another, and which each square to $+1$. Using those multiplication rules, we can furthermore generate on the one hand the scalar $1$, and on the other hand three basis \emph{bivectors} $\mathbf e_1 \mathbf e_2$, $\mathbf e_1 \mathbf e_3$, $\mathbf e_2 \mathbf e_3$ and a basis \emph{trivector} $\mathbf e_1 \mathbf e_2 \mathbf e_3$. Our algebra consists exactly of all real linear combinations of the scalar $1$, the three basis vectors, the three basis bivectors, and the single basis trivector. All this makes $\text{Cl}_{(3, 0)}(\mathbb R)$ a $1 + 3 + 3 + 1 = 8$ dimensional real vector space endowed with a compatible non-commutative but associative multiplication operation .

Real multiples of $1$ are called scalars, real linear combinations of $\mathbf e_1$, $\mathbf e_2$, and $\mathbf e_3$ are called vectors, real linear combinations of $\mathbf e_1 \mathbf e_2$, $\mathbf e_1 \mathbf e_3$, $\mathbf e_2 \mathbf e_3$ are called bivectors, and real multiples of $\mathbf e_1 \mathbf e_2 \mathbf e_3$ are called trivectors, and also called pseudo-scalars. The scalars $0$ and $1$ play, as elements of the algebra, the roles of an (additive and multiplicative) zero and of a multiplicative identity. Every element of the algebra can be written in a unique way as a sum of a scalar, vector, bivector and trivector. The algebra is called \emph{graded}; it is built up of elements of grades 0, 1, 2 and 3. We think of the vectors in the algebra as real 3D vectors in Euclidean space. The bivectors can be thought of as oriented plane elements, the trivectors as oriented volume elements. The bivectors together with the scalars form a four dimensional sub-algebra. It is the algebra of the quaternions, discovered by Hamilton.

I will start with some notational problems. The main mathematical problem will come later, and cannot be resolved by cleaning up the notation, as I will show. Ambiguous notation is just a warning signal.

Equations (32) and (33) of \cite{IEEEAccess2} introduce bivectors $\mathbf L(\mathbf a, \lambda)$, $\mathbf L(\mathbf b, \lambda)$, $\mathbf D(\mathbf a)$ and $\mathbf D(\mathbf b)$. Here, $\mathbf a$ and $\mathbf b$ are ordinary unit length 3D vectors, but also seen as elements of grade 1 in the Geometric Algebra. The second formal argument of $\mathbf L$ is the scalar called $\lambda$, which can take the values $+1$ and $ -1$. We are told in (32) that $\mathbf L(\mathbf a, \lambda) = \lambda \mathbf D(\mathbf a) = \lambda I  \mathbf a$ and in (33) that $\mathbf L(\mathbf b, \lambda) = \lambda \mathbf D(\mathbf b) = \lambda I  \mathbf b$ where $I$ is the trivector $\mathbf e_1  \mathbf e_2  \mathbf e_3$.  (Christian writes $\mathbf L(\mathbf a, \lambda) = \lambda I \cdot \mathbf a$ and  $I = \mathbf e_1  \wedge \mathbf e_2 \wedge \mathbf e_3$ but the symbols ``$\cdot$'' and ``$\wedge$'' in these contexts are superfluous). The pseudo-scalar $I$ commutes with everything and $I^2= -1$. The scalar $\lambda$ commutes with everything and $\lambda^2 = +1$.

It follows directly from Christian's (32) and (33)  that 
$$
\mathbf L(\mathbf a, \lambda) \mathbf L(\mathbf b, \lambda) ~=~ \lambda^2 I^2 \mathbf a \mathbf b ~=~ - \mathbf a \mathbf b,
\eqno(4)
$$ 
which does not depend on $\lambda$ at all. In fact, from geometric algebra we know that
$$
- \;\mathbf a\; \mathbf b ~=~ - \mathbf a \cdot \mathbf b - I \; \mathbf a\!\times\! \mathbf b ~=~ - \mathbf a \cdot \mathbf b - \mathbf L(\mathbf a \times \mathbf b, +1).
\eqno(5)
$$
But Christian's next equations (34) $\mathbf L(\mathbf a, +1)\mathbf L(\mathbf b, +1)=\mathbf D(\mathbf a)\mathbf D(\mathbf b)$ and (35)  $\mathbf L(\mathbf a, -1)\mathbf L(\mathbf b, -1)=\mathbf D(\mathbf b)\mathbf D(\mathbf a)$ could not then both be correct since $\mathbf D( \mathbf a)$ and  $\mathbf D( \mathbf b)$ do not commute, see his (36) $\mathbf D(\mathbf a)\mathbf D(\mathbf b) = -\mathbf a \cdot \mathbf b -\mathbf D(\mathbf a\times\mathbf b)$; the cross product of ordinary vectors does not commute. See also the in-line formula immediately before his (41), $\mathbf D(\mathbf n) = I \,\mathbf n$. 

Christian states, just after his (28), that the scalar $\lambda$ stands for the ``handedness'' of a basis of the tangent space at any point of $S^3$. The tangent space at every point is $\mathbb R^3$. If one fixes an orthonormal basis of the tangent space at one point, one can label its elements, and try to move it smoothly around the manifold. It is a fact that one can move around the manifold $S^3$ and discover that the labelling has changed when one comes back to the same position. One needs to complete the circuit twice to get back to the initial configuration. This is Dirac's famous belt trick, cf.\ Christian's Möbius strip example in his Section II. 

Is it possible that the contradiction just obtained follows from unfortunate notation? Perhaps the author has in mind both a right-handed and a left-handed cross product. And even, perhaps, a right- and left-handed geometric product? Can we restore consistency by introducing either of these features explicitly?

For example, consider what happens if we keep one Geometric Algebra product, but introduce two Euclidean space cross-products, $\times_{\!(\lambda)}$ where $\lambda =  \pm 1$, by the rules 
$$
\mathbf  a \times_{\!(+1)} \mathbf b ~=~ \mathbf a\! \times \!\mathbf b, \qquad \mathbf a \times_{\!(-1)} \mathbf b ~=~ \mathbf b \!\times\! \mathbf a.\
\eqno(6)
$$ 
Now equation (36), corrected, makes sense and is consistent with what follows:
$$
\mathbf L(\mathbf a, \lambda) \mathbf L(\mathbf b, \lambda) ~=~ - \mathbf a \cdot \mathbf b - \mathbf L(\mathbf a \times_{\!(\lambda)}\! \mathbf b, \lambda).
\eqno(7)
$$
Having restored consistency to the definitions we can now quickly check formulas (34) and (35), taking account of (36), which give us 
$$
\mathbf L(\mathbf a, \lambda) = \lambda I \mathbf a, \quad\mathbf D(\mathbf a) = \lambda \lambda I \mathbf a = I \mathbf a.
\eqno(8)
$$

However, there is no point at all in trying to fix these notational issues.  Let us jump to the definition of the important \emph{measurement functions}, $\mathcal A$ and $\mathcal B$. I will not use any modification of earlier definitions. I start with the left hand parts of Christian's defining equalities (39) and (40); the ones with limits as $\mathbf s_1$ converges to $\mathbf a$ and as $\mathbf s_2$ converges to $\mathbf b$ of a $\mathbf D$ times an $\mathbf L$. Those limits can be written down immediately, giving us 
$$
\mathcal A(\mathbf a, \lambda)= -\mathbf D(\mathbf a) \mathbf L (\mathbf a, \lambda) = - \lambda I^2 \mathbf a^2 = \lambda,
\eqno~~(9)
$$
$$
\mathcal B(\mathbf b, \lambda)= +\mathbf D(\mathbf b) \mathbf L (\mathbf b, \lambda) =  \lambda I^2 \mathbf b^2 =-\lambda,
\eqno(10)
$$
exactly as Christian himself tells us with the right hand sides of his (39) and (40), and consistent with his (43)--(47). Moreover, in (47) to (49), Christian writes explicitly that the product of $\mathcal A$ and $\mathcal B$ is identically equal to minus 1. 

For this very reason, Christian embarks on alternative but more complex computations of the same quantity, and arrives at a very different result, namely the actual singlet correlations. These alternative calculations must be wrong. Anyone who has some patience and can do elementary calculus will easily locate fatal errors. Christian attempts to evaluate the product of two limits, one as $\mathbf s_1$ converges to $\mathbf a$, the other as $\mathbf s_2$ converges to $\mathbf b$. He evaluates this limit by \emph{imposing} $\mathbf s_1 = \mathbf s_2 = \mathbf s$ \emph{before} taking the limit.  Taking the product first, the variables $\mathbf s_1$ and $\mathbf s_2$ happen to cancel, since $\mathbf s^2 = 1$. He now takes the limit as $\mathbf s_1\to\mathbf a, \mathbf s_2\to\mathbf b$ of an expression which does not depend on  $\mathbf s_1, \mathbf s_2$ at all.

Christian rounds things off with a computer simulation written in the language of the Geometric Algebra package \texttt{GAViewer}, which accompanied the book Dorst et al. (2007) \cite{dorst}. The package does not run on present day Mac or Linux machines without recompiling and building it from the source code. Fortunately, the code in Christian's paper is easy to read, with the help of the \texttt{GAViewer} user's manual.

The outcomes of the measurement functions are computed but ignored, except in order to compute their averages, which of course are close to zero. The program jumps to an intermediate step in Christian's theoretical evaluation of the product of the outcomes. Christian's programmer A. Wonninck here inserted the revealing line
$$
\texttt{if(lambda==1) \{q=A B;\} else \{q=B A;\}},
\eqno(11)
$$ 
which serves to switch between the geometric product and its transpose according to the sign of $\lambda$.  The correlations the program finds are not the averages of products of pairs of outcomes defined through Christian's measurement functions. The just mentioned program variables \texttt{A} and \texttt{B} are quaternions, $\texttt{A} = - I \mathbf a \lambda I \mathbf s_1 = \lambda \mathbf a \,\mathbf s_1$,  $\texttt{B} = \lambda I \mathbf s_2 I \mathbf b = - \lambda \,\mathbf s_2 \,\mathbf b$, where $\mathbf s_2 = \mathbf s_1 = \mathbf s$, say. Their product depends on the order of multiplication.  Since $\mathbf s ^2 = 1$, $\texttt{AB} =  - \mathbf a \mathbf b$, while $\mathtt{BA} = (-\mathbf s \mathbf b)( \mathbf a \mathbf s) =  -(- \mathbf b \mathbf s) (- \mathbf s \mathbf a) = - \mathbf b \mathbf a$. 

About half the time, the ``product of the measurements'' is \emph{defined by the code} as the quaternion $-\mathbf a \mathbf b$, the other half of the time it is the quaternion $- \mathbf b \mathbf a$. Recall the fundamental facts of 3D Geometric Algebra
$$
\mathbf a\: \mathbf b ~=~ \mathbf a\cdot\mathbf b + I\;\mathbf  a \!\times\! \mathbf b,
$$ 
$$
{\textstyle\frac 12}\;\mathbf a\;\mathbf b + {\textstyle\frac 12}\; \mathbf b\; \mathbf a ~=~ \mathbf a\cdot\mathbf b.
$$
The program randomly samples many uniformly distributed, independent unit length vectors $\mathbf a$ and $\mathbf b$. For a given pair, it computes either  $-\mathbf a \mathbf b$ or  $-\mathbf b \mathbf a$, chosen by the outcome of a fair coin toss. It also computes the arc cosine of $\mathbf a\cdot\mathbf b$. The scalar part of the average of the geometric products is plotted against the angles $\cos^{-1}(\mathbf a\cdot\mathbf b)$, grouped into bins. The bivector parts are printed to show they are small, but they have to be discarded from the plot anyway.

Christian's companion \emph{IEEE Access} paper \cite{IEEEAccess1} also concluded with a computer simulation. His computer programmer for that paper, C.F.~Diether, adopted Gill's \cite{Gill2020c} implementation of Pearle's (1970) \cite{Pearle} detection loophole model. Gill had fixed errors in Pearle's (1970) classic paper, posted \texttt{R} code on internet, and discussed it in public internet discussions in early 2014. That was the first time that anyone had implemented the Pearle model as a computer simulation. Till then, it had been seen as a purely theoretical result on the minimal efficiency of detectors needed to violate the CHSH inequality in the face of the detection loophole when using the usual state and spin measurements. Gill (2020a) \cite{Gill2020a} already showed that \cite{IEEEAccess1} contains the same defects as \cite{IEEEAccess2}.

\section{Conclusion}

Christian's model simply reproduces the Bertlmann effect. Bertlmann always wore one pink and one blue sock, at random. The moment you saw his left foot, you knew what sock he wore on his right foot. As John Bell explained \cite{Bertlmann}, quantum entanglement \emph{cannot} be explained in such a way. Christian's model assigns the two particles of the EPR-B experiment an equal and opposite spin at the source, and however they are measured, these spins are recovered.

\section{Appendix}

\subsection{Inside Bell's theorem: no-go theorems in probability theory and in distributed computing}
I see Bell's theorem as the metaphysical statement that quantum mechanics (QM) is incompatible with local realism (LR). More precisely, and following Tsirelson's \texttt{Citizendium.org} article \cite{Tsirelson}, Bell's theorem states that conventional quantum mechanics is a mathematical structure incompatible with the conjunction of three mathematical properties: \emph{relativistic local causality} (commonly abbreviated to ``locality''), \emph{counterfactual definiteness} (``realism'') and \emph{no-conspiracy} (``freedom''). By conventional quantum mechanics, I mean: quantum mechanics including the Born rule, but with a minimum of further interpretational baggage. Whether the physicist likes to think of probabilities in a Bayesian or in a frequentist sense is up to them. In Many Worlds interpretations (and some other approaches), the Born rule is argued to follow from the deterministic (unitary evolution) part of the theory. But anyway, everyone agrees that it is there.

Bell himself, a physicist writing for physicists, sometimes used the phrase ``my theorem'' to refer to his \emph{inequalities}: first his (Bell, 1964) three correlations inequality \cite{Bell}, and later what is now called the Bell-CHSH (Clauser, Horne, Shimony, Holt 1969) \cite{CHSH} four correlations inequality (2). I see those inequalities as simple probabilistic \emph{lemmas} used in Bell's various proofs over the years of the same \emph{theorem} (the incompatibility of QM and LR). 

Fine (1982) \cite{Fine} converted Bell's theorem into an ``if and only if result''. He showed that the satisfaction of all eight one-sided Bell-CHSH inequalities together with the four no-signalling equalities is necessary and sufficient for a local hidden variables theory to explain the sixteen conditional probabilities $p(x, y\mid a, b)$ of pairs of binary outcomes $x$, $y$ given pairs of binary settings $a$, $b$, in a Bell-CHSH-type experiment. (No-signalling is the statement that Alice can not see from her statistics, what Bob is doing:  $p(x\mid a, b)$ does not depend on $b$, and similarly for Bob, $p(y\mid a, b)$ does not depend on $a$.) A precursor of Fine's theorem can be recognised in Boole's (1853) book \cite{Boole}. Illustrating general methodology developed in his book, Boole derives the conditions on three probabilities $p$, $q$ and $r$ of three events which must hold in order that a probability space exists on which those three events can be defined with precisely those three probabilities, given certain logical relations between those three events, and comes up with what can be recognised, with some creativity, as the six one-sided Bell three-correlation inequalities. With four events, his methodology would have given us Fine's theorem. More recently these results have been generalised to experiments with arbitrary numbers of parties, measurement settings, and measurement outcomes, \cite{Brunner}.

In a Bell-CHSH type experiment we have two locations or labs, in which two experimenters Alice and Bob can each choose a binary setting to a device, which then generates a binary output or measurement outcome. The experimenters have previously set things up so that the setting choices correspond to certain angles or directions. Just two settings are considered in each wing of the experiment. This is repeated, say $N$ times. We will talk about one \emph{run} consisting of $N$ individual \emph{trials}. The binary setting choices are externally generated, perhaps by tossing coins or performing some other auxiliary experiment. One published experiment used, as inputs, the bits of a maximally compressed video recording of the movie ``Back to the Future''. The spatial-temporal arrangement of the $2N$ measurements is such that there is no way a signal carrying Alice's $n$th setting, sent just before it is inserted into her device, could reach Bob's lab before his device has generated its $n$th outcome, even if transmitted at the speed of light, and vice versa.

These experiments involve measurements of the ``spin'' of ``quantum spin-half particles'' (electrons, for instance); or alternatively, measurements of the polarization of photons in the plane opposite to their directions of travel. The two settings, both of Alice and Bob, correspond to two \emph{directions} (spin) or \emph{orientations} (polarization), usually in the plane, but conceivably in three-dimensional space. (Polarization can be represented as a direction in 3D, on what is called the Poincaré sphere). Focussing on the case of spins: in the so-called singlet state of two entangled spin-half quantum systems, one can conceivably measure each subsystem in any 3D direction whatsoever, and the resulting pair of $\pm 1$-valued outcomes $(X_{\mathbf a}, Y_{\mathbf b})$ would have the ``correlation'' $\mathbb E X_{\mathbf a} Y_{\mathbf b} = - {\mathbf a} \cdot {\mathbf b}$. Marginally, they would be completely random, $\mathbb E X_{\mathbf a}= \mathbb E Y_{\mathbf b} = 0$.

These statistical predictions are easy to compute, using the standard rules of quantum mechanics, for the so-called EPR-B experiment: the Einstein, Podolsky, Rosen (1935) thought experiment, transferred to spin by Bohm and Aharonov (1957). (Translated to the polarization example, this joint probability distribution of two binary variables is often called Malus' law.) We will stick to the spin-half terminology and talk about ``the singlet correlations'' referring to the whole family, indexed by pairs of directions, of joint probability distributions of two binary variables just described. The archetypical example (though itself only a thought experiment) of such an experiment would involve two Stern-Gerlach devices and is a basic example in many quantum physics texts.

Bell was interested in what one nowadays calls (stochastic) ``local hidden variables theories'' (LHV).  According to such a theory, the statistics predicted by quantum mechanics, and observed in experiments, are merely the reflection of a the classical underlying theory of an essentially deterministic and local nature. There might be local randomness, for instance, further randomness in the measurement devices. Different sources of randomness could even be correlated. Mathematically, such theories are generally agreed to assert the mathematical existence of a classical probability space on which are defined random variables $X_{\mathbf a}$ and $Y_{\mathbf b}$ for all directions ${\mathbf a}$ and ${\mathbf b}$ in the plane (or in space), such that each pair $(X_{\mathbf a}, Y_{\mathbf b})$ has got the previously described joint probability distribution. A natural mathematical question is: can such a probability space exist? The answer is well-known to be ``no''. Just one of the many ways to prove this theorem is through Bell's inequalities.

The underlying probability space is usually called $\Lambda$ instead of $\Omega$, and the elementary outcomes $\lambda\in\Lambda$ stand for the configuration of all the particles involved in the whole combined set-up of a source connected to two distant detectors, which are fed the settings ${\mathbf a}$ and ${\mathbf b}$ from outside. Thus $X_{\mathbf a}(\lambda)$ stands \emph{within the mathematical model} for the outcome which Alice would theoretically see if she used the setting ${\mathbf a}$, even if she actually used another. There is no claim that these variables exist in reality, whatever that means. We are talking about the mathematical existence of a model with certain mathematical properties.

In a sequence of trials, one might initially suppose that for each trial there is some kind of resetting of apparatus, so that at the $n$th trial we see the outcomes corresponding to $\lambda = \lambda_n$, where the sequence $\lambda_1$, $\lambda_2$, \dots, are independent draws from the same probability measure on the same probability space $\Lambda$. Now suppose we could come up with such a theory, and indeed come up with a (classical) Monte-Carlo computer simulation of that theory, on a classical PC. Then we could do the following. Simulate $N$ outcomes of the hidden variable $\lambda$, and simply write them into two computer programs as $N$ constants defined in the preamble to the programs. More conveniently, if they were simulated by a pseudo random number generator (RNG), then we could write the constants used in the generator, and an initial seed, as just a few constants, and reproduce the RNG itself inside both programs. The programs are to be run on two computers thought of as belonging to Alice and Bob. Think of the case of directions in the plane. The two programs are started. They both set up a dialogue (a loop). Initially, $n$ is set to $1$.  Alice's computer prints the message ``Alice, this is trial number $n=\ldots$. Please input an angle.'' Alice's computer then waits for Alice to type an angle and hit the ``enter'' key. Bob's computer does exactly the same thing, repeatedly asking Bob for an angle.

If, on her $n$th trial, Alice submits the direction ${\mathbf a}$, then the program on her computer evaluates and outputs $X_{\mathbf a}(\lambda_n)=\pm 1$, increments $n$ by one, and the dialogue is repeated. Alice's computer does not need Bob's direction for this -- locality! Bob's does not need Alice's. Thus, if one could implement a local hidden variables theory for \emph{one trial} of a Bell-type experiment in \emph{one} computer program, then one could simulate the singlet correlations derived from one run of many trials on \emph{two completely separate} computers, each running its own program, and each receiving its own stream of inputs (settings) and generating its own stream of outputs.

This idea goes back a long way. It is for instance mentioned in Jaynes (1999) \cite{jaynes}. In his paper, presented at the MaxEnt conference the preceding year, Jaynes had argued that Bell did not understand conditional probability. The paper was discussed by Steve Gull, who disagreed, and had posed the problem: ``Write a program which is to run on two PC's which mimics the QM predictions for the EPR setup. There must be no communication between the computers after the time of program load''. He then presented a ``Sketch proof of impossibility'' using Fourier theory, no Bell inequalities at all. Jaynes was dumbfounded and predicted that it would take 30 years to understand Gull's new result, just as it had taken 20 years to understand Bell's. Gull's overhead transparencies are reproduced \cite{gull} on his home page. Gill and Karakozak (2020) \cite{GillKarakozak} have worked out the proof in full detail.

But why should those two functions be the same, for each trial? Maybe the parameters of the underlying physics drifts, and even occasionally jumps. Experimenters know there is an enormous stability problem in these experiments. The LHV model for the $n$th trial should be allowed to depend on all previous inputs and outputs of all previous trials, as well as depending on time, $n$. The only thing that is forbidden, is that Alice's $n$th output depends on Bob's $n$th input, and vice versa.

Focussed on classical networked computer simulations of stringent CHSH type experiments, Gill (2003) \cite{Gill2003} produced a martingale inequality using a variant of the CHSH statistic. He considered the case that settings are chosen completely at random. He noticed that if the denominators of the four fractions defining the four sample correlations are replaced by their expectation values $N/4$ and the whole statistic is multiplied by $N/4$, it then equals a sum over the $N$ trials of a quantity which, under local realism, and assuming the complete randomness of the \emph{setting} choices only, has conditional expectation (given the past) less than $3/4$. Subtract off $(3/4)N$ and one has, under local realism, a supermartingale in the time variable $N$. The conditional expectation of increments of the process, conditional on the past, are negative. The increments of the process are bounded, and martingale theory supplies powerful exponential inequalities on the probabilities of large deviations upwards.

Gill's methods were further refined. Hensen et al. (2015) \cite{Hensen2015}, reporting the first ever successful ``loophole-free'' Bell-type experiment in the journal \emph{Nature}, also derived and employed a new martingale based modification of the CHSH inequality. Consider such an experiment and let us say that the $n$th trial results in a success, if and only if the two outcomes are equal and the settings are the pair $(1, 2)$, or the two outcomes are opposite and the settings are not the pair $(1, 2)$. The quantum engineering is such as to ensure a large positive correlation between the outcomes for setting pair $(1,2)$ and a large negative correlation for setting pairs $(1,1)$, $(2,1)$ and $(2,2)$. Let us denote the total number of successes in a fixed number, $N$, of trials, by $S_N$. Then Hensen et al.~(2015) \cite{Hensen2015} show that, for all $x$, under the assumption of local realism,
$$
\mathbb P(S_N\ge x)~\le~ \mathbb P( \mathrm{Bin}(N, 3/4) \ge x),
\eqno(12)
$$
where $\mathrm{Bin}(N, p)$ denotes a binomially distributed random variable with parameters $N$, the number of trials, and success probability per independent trial $p$.

``The assumption of local realism'' is a bundle of physics concepts. Notice that a network of two classical PC's both performing a completely deterministic computation, and allowed to communicate over a classical wired connection \emph{between} every trial and the next, but not during each trial, does satisfy those assumptions. The theorem applies to a classical distributed computer simulation of the usual quantum optics lab experiment. Time trends and time jumps in the simulated physics, and correlations (dependency) due to use of memory of past settings (even of the past settings in the other wing in the experiment) do not destroy the theorem. The probability inequality (12) is driven \emph{solely} by the completely random choice anew, trial after trial, of one of the four pairs of settings, while each computer is only fed its own setting, not that given to the other computer. Statistical randomisation neutralises effects of uncontrolled (and maybe even unknown) confounders, and leads to guaranteed $P$-values.

Take for instance $N = 10\,000$. Take a critical level of $x = 0.8 N$. Local realism says that $S_N$ is stochastically smaller (in the right tail) than the  $\mathrm{Bin}(N, 0.75)$ distribution. According to quantum mechanics, and using the optimal pairs of settings and the optimal quantum state, $S_N$ would have the $\mathrm{Bin}(N, p)$ distribution with $p = \frac12 + \frac{\sqrt 2}4 \approx 0.85$. Under those two distributions, the probabilities of outcomes respectively larger and smaller than $0.80 N$ are about $10^{-30}$ and $10^{-40}$ respectively. I challenge anyone who believes in local realism to develop a computer simulation of a local realistic physical model, which subject to the external experimental constraints sketched by Bell in \cite{Bertlmann}, and nowadays routinely imposed in ``loophole free Bell experiments'', reliably achieves a greater than 80\% large $N$ success rate. The simulation experiment must be reproducible through the use of a ``set seed'' facility in any used RNG, so that it can be verified by intensive testing that Alice's $n$th output does not depend on Bob's $n$th input, nor on future inputs of Alice or Bob. The inputs must not be generated inside the simulation, but must be supplied by the outside user. The statistical analysis of the outputs must also be left to the user.

\subsection{Metaphysics: what should we believe, now?}

Several reviewers preferred novel physics and metaphysics in a paper on Bell's theorem, instead of well known mathematics. I have the same preference but as a mathematician, not a physicist or a philosopher, I am not qualified to deliver. My ``position'' on the metaphysical or philosophical issues has varied over the years and remains open. I see several reasonable positions to hold. I do think that since the loophole-free experiments of 2015, ``local realism'' is no longer tenable. Those experiments do need improvement. For instance, Hensen et al.'s (2015) Delft experiment \cite{Hensen2015} had $N=245$, far too small. The $P$-value for testing the null-hypothesis of local realism was $3\%$. The result was promising and the experimental set-up was brilliant and innovative (making use of a technique called entanglement swapping), and moreover a model of ``good scientific practice''.

\begin{IEEEbiography}[{\includegraphics[width=1in,height=1.25in,clip,keepaspectratio]{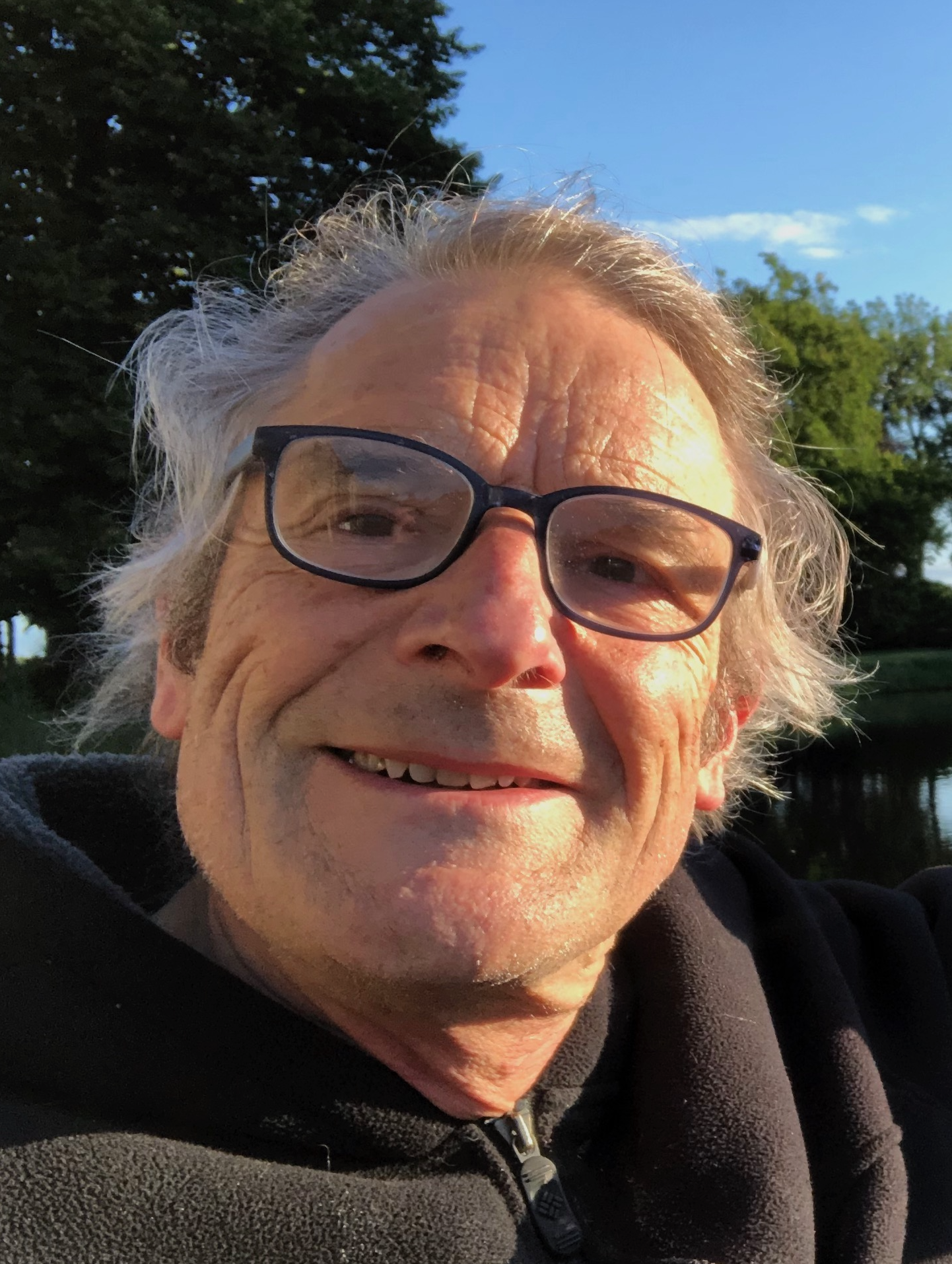}}]{Richard D. Gill} 
was born in 1951 in the UK. B.A. degree in mathematics, Cambridge University, 1973; diploma of statistics, Cambridge University, 1974; PhD degree in mathematics, Free University Amsterdam, 1979. In his career he has been head of statistics department, CWI Amsterdam; professor mathematical statistics in Utrecht, later in Leiden; and is now emeritus professor in Leiden. His early work was in counting processes, survival analysis, martingale methods, semiparametric models. Later he has worked in forensic statistics, quantum information, and on scientific integrity. His work on experimental loopholes in Bell-type experiments was incorporated in the famous ``loophole-free'' Bell experiments of 2015. He is a member of the Royal Dutch Academy of Science and a past president of the Netherlands Society for Statistics and Operations Research.
\end{IEEEbiography}

\EOD

\end{document}